\documentclass[
 reprint,
 amsmath,amssymb,aip
]{revtex4-1}

\newcommand{\vt}{v_{T_e}}

\newcommand{\erf}{\textrm{erf}}

\usepackage{graphicx}% Include figure files
\usepackage{dcolumn}% Align table columns on decimal point
\usepackage{bm}
\usepackage{tikz}
\usetikzlibrary{shapes,arrows}
\begin{document}

%\preprint{APS/123-QED}

\title{Presheath-like structures and effusive  particle losses for biased probes at and near I-V  electron saturation   }% Force line breaks with \\

\author{Brett Scheiner}
 \affiliation{Lam Research Corporation, Tualatin OR 97062}%Lines break automatically or can be forced with \\

\date{\today}% It is always \today, today,

\begin{abstract}

A theory for presheath-like structures near probes biased at and above the plasma potential is developed for collisionless plasmas with an electron-neutral mean free path on the order of the chamber scale. 
The theory predicts presheath-like perturbations to the plasma that result from the free streaming of electrons and an effusion loss process from the chamber at the electrode. For these situations, a loss-cone-like velocity distribution function for electrons is predicted where the loss angle of the depletion region corresponds to the angular size of the electrode at a specified distance. The angle of the loss cone becomes 180 degrees at the sheath edge.
In comparison to a previous collisional electron presheath model that required electrons satisfy a Bohm criterion at the sheath edge [B. Scheiner et al. Phys. Plasmas 22, 123520 (2015)], the present work suggests that no such condition is needed for collisionless low pressure plasmas in the $\lesssim$10 mTorr range. 
The theory predicts the generation of a density depletion of roughly 0.5$n_e$ and an electron velocity moment of 10s of percent of the electron thermal speed by the sheath edge in a presheath with a potential drop of less than $T_i/e$. 
The range of this presheath perturbation is determined by the electrode geometry instead of the collisional mean free path. These predictions are tested against previously published particle in cell simulations and are found to be in good agreement.   

\end{abstract}

\maketitle

%\tableofcontents

\section{Introduction}

\begin{figure}[h]
    \centering
    \includegraphics[width=0.45\textwidth]{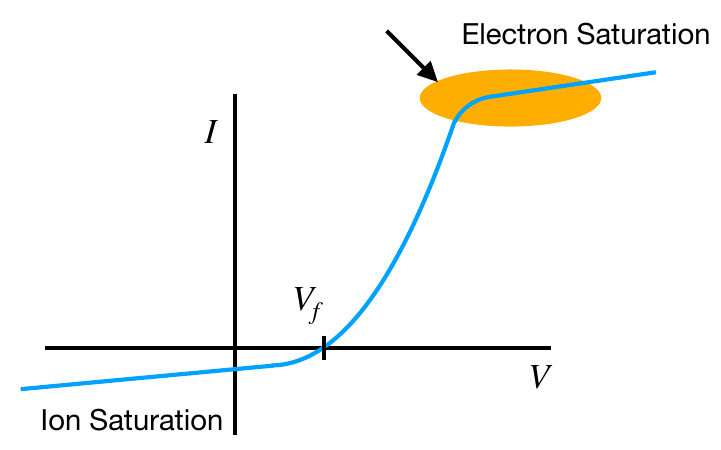}
    \caption{A sketch of a Langmuir probe current voltage (I-V) trace. The location of the floating potential and electron and ion current saturation are indicated. The orange highlighted area indicates the area of study for this paper. }
    \label{fig:IV}
\end{figure}

Langmuir probes are the most common and oldest diagnostic for laboratory plasma experiments\cite{PhysRev.28.727}. Their operation is simple: sweep the DC bias of a probe such as a conducting wire or disk electrode and measure the resulting current. The plasma properties inferred from the current-voltage response (c.f. Fig.~\ref{fig:IV}) depend on a theory of how the probe collects ions and retards electrons. Such theory allows one to determine the density, temperature, and electron velocity distribution function (EVDF). The most widely used theories describe the probe-plasma interaction in the electron retarding regime 
on the basis of orbital motion in a central potential\cite{PhysRev.28.727}, electron and ion contribution to space charge in the probe-plasma system\cite{JEAllen_1957}, or both combined\cite{10.1063/1.1705900,laframboise1966theory}, 
and have been discussed extensively\cite{electricprobes}. 
%Briefly summarizing, the most well known and widely applied theory is Orbital Motion Limited (OML) theory\cite{PhysRev.28.727} which considered the orbital motion and conservation of angular momentum and energy of collected ions. OML theory assumed a finite absorption radius and due to this limitation Allen-Boyd-Reynolds (ABR) theory considered the Poisson solution in all space allowing the calculation to consider particles collected at infinity\cite{JEAllen_1957}. However, all ions in ABR theory are assumed to be collected by the probe and orbital motion is not considered. Bernstein-Rabinowitz-Laframboise theory improved upon this by considered both the space potential and the orbital motion of ions\cite{10.1063/1.1705900,laframboise1966theory}. 
Since most information is gleaned from Langmuir probes operating in this regime little attention has been given to the electron saturation regime (Ref.~\onlinecite{10.1063/1.1728342} is one notable exception). This leaves an open question of whether or not useful information can be extracted from the probes under these conditions. This work explores the mechanism of electron collection and the perturbation imposed on the bulk plasma by an electrode in the electron saturation regime in low pressure collisionless plasmas. In doing so we show that the plasma-probe interface is fundamentally different from that which is encountered when the probe is electron retarding or in ion saturation.

Electron saturation, depicted in the Langmuir probe I-V trace sketch of Fig.~\ref{fig:IV}, occurs when the probe is biased positive relative to the plasma potential such that it is electron collecting. 
Probes at electron saturation have an electron sheath, an electron rich sheath whose potential increases from the plasma to the probe surface. 
Until the past decade it was typically assumed that the electron sheath posed little perturbation to the bulk plasma. However, a series of experimental and simulation results published by myself and coauthors showed that at least in some cases the electron sheath is accompanied by a presheath structure that extends into the bulk plasma over a length scale much longer than that of the ion sheath\cite{Yee_2017,10.1063/1.4939024,10.1063/1.4960382,Baalrud_2020}. 
These measurements used laser collision induced fluorescence to measure 2D profiles of the resulting bulk plasma electron density perturbation in the vicinity of an electron sheath in a 20 mTorr helium plasma\cite{Yee_2017}.
A fluid model developed to explain these observations described the electron presheath as a pressure gradient driven flow of electrons towards the electrode. Other observations predicted by the model and borne out in simulation data included the absence of the $T_e/e$ presheath potential drop seen in the ion presheath\cite{Yee_2017,10.1063/1.4939024}, a presheath electron flow velocity with magnitude near the electron thermal speed\cite{Yee_2017,10.1063/1.4939024,10.1063/1.4960382}, and the presence of instabilities\cite{10.1063/1.4939024} which recently have been observed in experiments\cite{10.1063/1.5142014}.

The most significant prediction of the electron sheath theory was that of an electron sheath Bohm criterion that reguired $V_e\ge\sqrt{(T_e+T_i)/m_e}$ by the sheath edge\cite{Yee_2017,10.1063/1.4939024}. This prediction was based on the premise that the EVDF in the presheath could be approximated as a Maxwellian with a flow shift. However maintenance of a Maxwellian EVDF with a flow shift requires that the presheath be collisional and that its length is significantly longer than the electron mean free path. At the 20 mTorr condition the previous experiments, the observed presheath was similar in length to the $\sim2$ cm mean free path. Furthermore, recent experiments at pressures of $\lesssim1$ mTorr measured electron sheath electron fluxes that are consistent with the collection of a random flux and no electron sheath Bohm criterion was indicated\cite{Jin_2022}. This raises the question of what the proper description is for electron sheaths in low pressure collisionless plasmas.

\begin{figure}[h]
    \centering
    \includegraphics[width=0.45\textwidth]{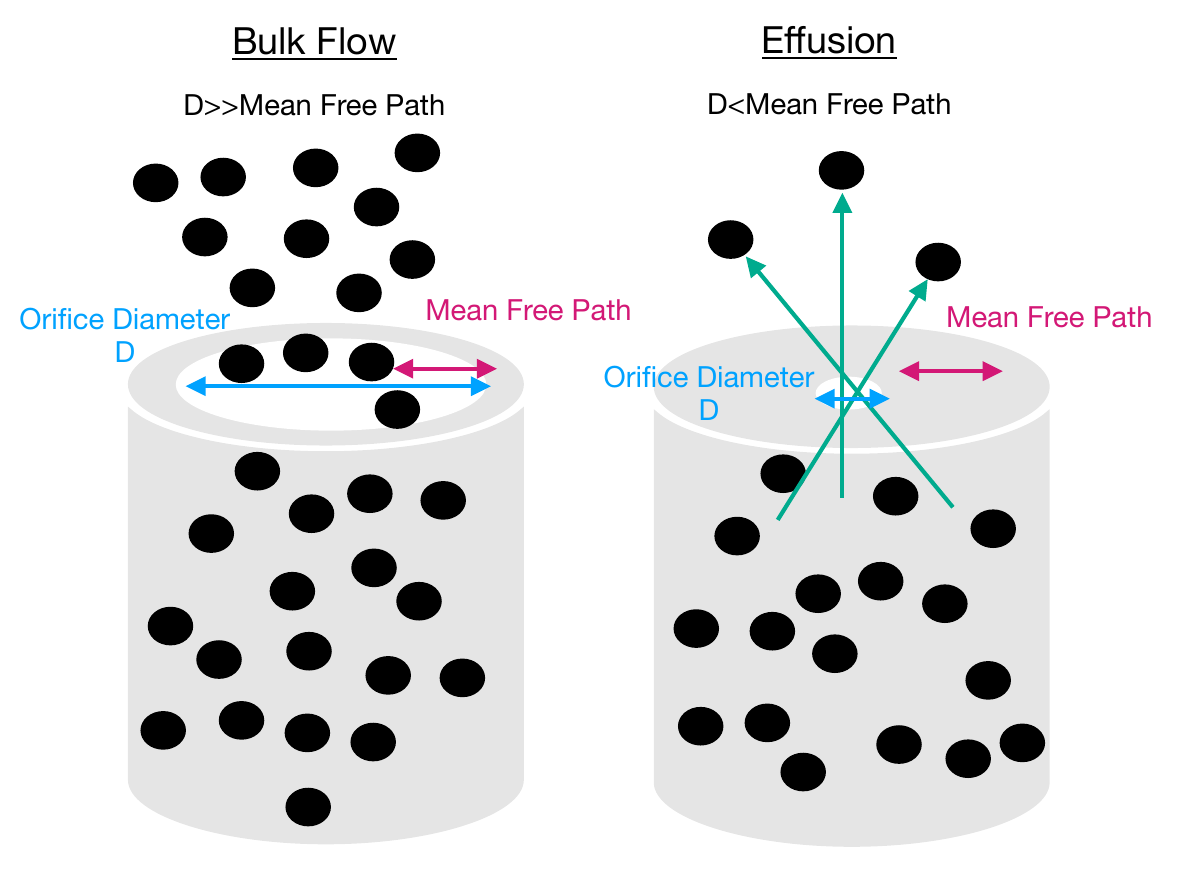}
    \caption{A comparison of bulk flow and effusion based losses from a gas container into vacuum. When the orifice through which the gas is lost is larger than the collisional mean free path the loss can be described by a pressure gradient driven bulk flow. Particles are lost by effusion when the orifice is smaller than the mean free path. For this case of only particles with free streaming trajectories intercepting the location of the orifice are lost. }
    \label{fig:Effusion}
\end{figure}

One peculiar observation from collisionless particle-in-cell (PIC) simulations was the formation of a loss cone distribution in the EVDF with the loss region oriented along the normal direction away from the electrode. The angle of this cone decreased, nearly closing to zero, while moving away from the electrode\cite{10.1063/1.4960382}. This observation along with the long mean free path of electrons suggest effusion as a mechanism for presheath formation. 
Effusion (c.f. Fig.~\ref{fig:Effusion}) is the process by which gas escapes from a container to vacuum through an orifice whose physical dimension is much less than the collisional mean free path. In such cases, particles with free streaming trajectories that pass through the orifice are lost. 
Contrasting with effusion, when the orifice is much larger than the collisional mean free path the loss from the container to vacuum can be described as bulk fluid flow driven by a pressure gradient. This paper describes the perturbation posed by probes at and near the electron saturation regime as an effusion process.  

This paper is organized as follows: Section~\ref{sec:theory} presents a theory for effusion based losses starting with the case of a probe biased exactly at the plasma potential in Sec.~\ref{sec:pp}. This serves as a starting point for the cases of probes biased slightly above and below the plasma potential (Sec.~\ref{sec:nearpp}), and the case of the electron presheath at probes biased above the plasma potential(Sec.~\ref{sec:epre}). In all cases, presheath-like structures are predicted even in cases where the absence of a sheath is predicted. Section III tests the theory against data from previously published 2D PIC simulations. The results are discussed in Sec. IV and a conclusion is given in Sec. V. 

\section{Theory\label{sec:theory}}
In this section a theory for boundaries with effusion electron losses is developed. Section~\ref{sec:pp} addresses the case where electrons are lost by effusion from a boundary biased exactly at the plasma potential. The theory predicts the EVDF and electron density and flow velocity profiles as a function of position and electrode size. This case serves as a starting point for understanding the behavior throughout the remainder of the paper. 
Section~\ref{sec:nearpp} discusses specific cases where the electrode is biased less than $\sim T_i/2e$ above or $\sim T_e/2e$ below the plasma potential. Section~\ref{sec:epre} considers applications to the electron presheath.    

\subsection{Boundaries at the plasma potential \label{sec:pp}}

We start by considering the form of the plasma-probe boundary when the probe is biased exactly at the plasma potential. Here, it is assumed that no retarding potential is present at the probe for electrons or ions. 
The theory developed in this section concerns plasmas that are collisionless to the extent that the electron-neutral and electron-electron mean free paths are longer than the length scales of the electron density and flow velocity profiles that are predicted by the theory.
It is assumed that particle trajectories are unaffected by any macroscopic electric field and that electrons and ions free stream until they impact the probe boundary. Interaction with other bounding sheaths of the plasma are not considered. 
%The only plasma boundary considered is the electrode.
%In consideration of these limitations, the theory generally applies to low temperature laboratory plasmas with neutral gas pressures less than about 10 mTorr.   

\begin{figure}[h]
    \centering
    \includegraphics[width=0.5\textwidth]{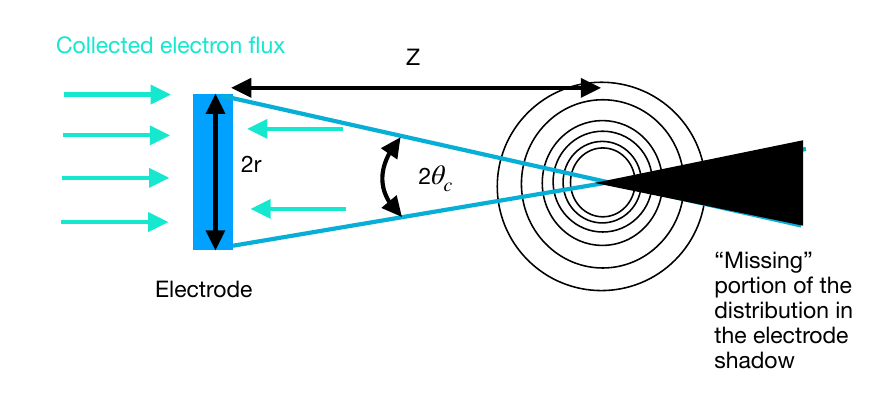}
    \caption{A sketch of the distribution function in relation to a disk electrode of radius r at distance z. The "missing" loss cone portion of the distribution corresponds to angles at which electrons are absent due to collection by the electrode.}
    \label{fig:shadow}
\end{figure}

Start with the assumption that electrons move along straight line trajectories until they intercept a conducting boundary at which they are lost from the system.
This behavior is analogous to the effusion through the circular orifice depicted in Fig.~\ref{fig:Effusion}. For the purpose of the present discussion we assume a disk shaped electrode with diameter $D=2r$.
As shown in Fig.~\ref{fig:shadow}, if one chooses a point some distance $z$ away from the disk along a line perpendicular to it's center, only electrons with velocity magnitude directed within the angle
\begin{equation}\label{eq:thetac}
    \theta_c(z)=\arctan(r/z)
\end{equation}
of the line (oriented in the $-\hat{z}$ direction) are collected. Likewise, assuming symmetry between the plasmas on the front and back side of the disk, electrons collected on the backside need to be within an angle of the same magnitude. 
Electrons with negative z velocity that are not collected move from the plasma on the front side to that on the back, and vice versa, such that electrons in the front side plasma with a velocity magnitude oriented in the positive z direction originated from the back-side plasma. 
Due to this fact, it is expected that the plasma at point z in front of the electrode is absent of electrons with velocities that are directed within $\pm\theta_c$ of the $\hat{z}$ direction. In this situation, the EVDF is determined geometrically and is     
\begin{equation}\label{eq:vdf}
f_e(z)=\frac{n_o}{\pi^{3/2}\vt^3}\exp\big(-v^2/\vt^2\big)H\big(\theta-\theta_c(z)\big).
\end{equation}
Here, $n_o$ is the electron density in the plasma as $z\to\infty$ such that $\theta_c\to 0$, $\vt$ is the electron thermal speed, $H(x)$ is the Heaviside function, and $\theta\in[0,\pi]$. 
It is worthwhile to note that the arguments above also apply to a position along a non-perpendicular line through the center of an electrode. In this case the electrode radius would be reduced by a factor of $\cos\alpha$, where $\alpha$ is the angle between the normal and the line in consideration. 

\begin{figure}[h]
    \centering
    \includegraphics[width=0.5\textwidth]{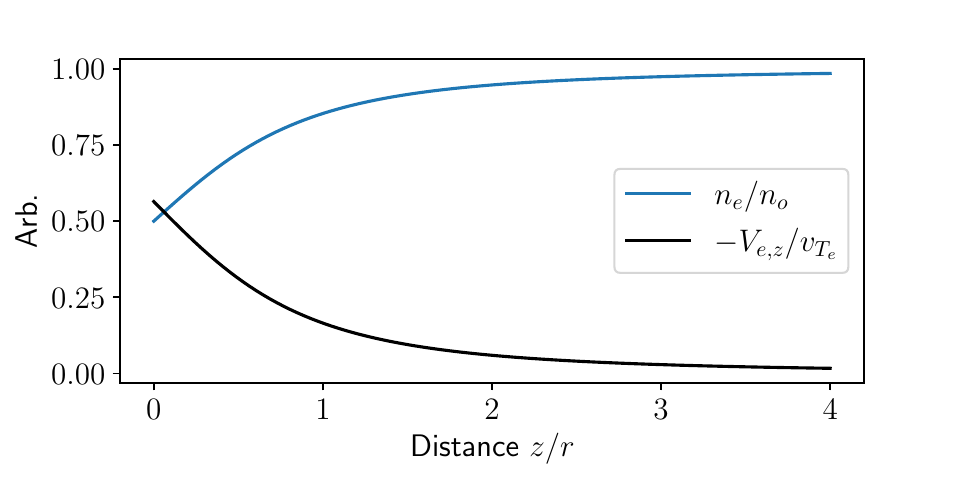}
    \caption{Profiles of electron density and flow velocity along the axis perpendicular to the electrode surface for the case of a disk electrode biased at the plasma potential. The scale length of the density and flow velocity are determined by the electrode radius $r$.}
    \label{fig:profile}
\end{figure}

The electron density and z flow velocity profiles can be determined by taking the $1$ and $v\cos\theta$ moments of the distribution function in Eq.~\ref{eq:vdf}. These are
\begin{equation}\label{eq:ne}
n_e=\frac{n_o}{2}(1+\cos\theta_c)
\end{equation}
and
\begin{equation}\label{eq:ve}
V_{e,z}=-\frac{\vt}{\sqrt{\pi}}(1-\cos\theta_c).
\end{equation}
The profiles of electron density and flow velocity corresponding to these profiles are shown in Fig.~\ref{fig:profile}. The key feature of these profiles is that the scale length of the perturbation imposed by the probe upon the bulk plasma is comparable to and dependent on the dimension of the electrode disk radius. For the case of the disk electrode this significantly exceeds the length scale of the sheath in many laboratory plasmas for a centimeter scale radius. For reference, the Debye length for a plasma with $T_e=2$ eV and $n_e=10^9/\textrm{cm}^2$ is 0.033 cm. Electrodes biased at or above the plasma potential that are significantly larger than this scale are expected to perturb the surrounding plasma. 

So far, only properties of electrons have been considered. 
For the present case of a probe biased at the plasma potential, one would expect the ion density to also be described by Eq.~\ref{eq:ne}. On the other hand, the ion velocity would be 
\begin{equation}\label{eq:vi}
V_{i,z}=-\frac{v_{T_i}}{\sqrt{\pi}}(1-\cos\theta_c).
\end{equation}
Since $n_i=n_e$, the plasma remains quasi-neutral to the boundary, despite the density rarefaction. However, Eqs.~\ref{eq:ve} and \ref{eq:vi} imply that the ratio of the electron to ion flux at each point in the profile including the probe surface is $\sqrt{T_e m_i/(T_i m_e)}$, just as it is through any plane defined in the bulk plasma in the absence of a boundary. If a sheath is not present to balance these loses, how can this be? The answer lies in the requirement that a probe be small compared to the other loss areas of the plasma (e.g. chamber walls) for it to be biased at or above the plasma potential, allowing it to collect more electrons than ions. If the probe is too large the plasma potential will always be above that of the surface\cite{10.1063/1.2722262}.

\subsection{Biases near the plasma potential\label{sec:nearpp}}

In this subsection, the situation where the bias is near but not equal to the plasma potential is considered. 
First we start with the case where the electrode bias is below the plasma potential by an amount that is not more than $\sim T_e/e$. 
This criteria is chosen such that there is a retarding potential for electrons, but not to an extent that nearly all electrode directed electrons are repelled. 
The validity of this restriction will be evaluated a posteriori. In this case the electron number density is made up of two components. The first term is the loss cone depleted distribution described by Eq.~\ref{eq:ne} while the second is the portion of electrons within the loss cone ($\theta<\theta_c$) that are repelled by the potential. These components are   
\begin{align}\label{eq:neR}
n_e=&\frac{n_o}{\pi^{3/2}\vt^3}\bigg[\int_0^{2\pi}d\phi\int_{\theta_c}^\pi d\theta \sin\theta \int_0^\infty dv' v'^2e^{-v'^2/\vt^2}\\
&+\int_0^{2\pi}d\phi\int^{\theta_c}_0 d\theta \sin\theta\int_0^{v_{c}} dv' v'^2e^{-v'^2/\vt^2}\bigg],\nonumber
\end{align}
where $v_c=\sqrt{2e|\Delta\phi_{R}|/m_e}$ is the velocity at which an electron has energy equal to the retarding potential $\Delta\phi_{R}$. The full result is 
%\begin{equation}
% 2\pi\frac{\vt^3\sqrt{\pi}}{4}\bigg[\erf\big(\frac{v_c}{\vt}\big)-\frac{2}{\sqrt{\pi}}\frac{v_c}{\vt}e^{-v_c^2/\vt^2}\bigg](1-\cos\theta_c)
%\end{equation}
\begin{align}
n_e=&\frac{n_o}{2}(1+\cos\theta_c)+\\
&\frac{n_o}{2}\bigg[\erf\big(\frac{v_c}{\vt}\big)-\frac{2}{\sqrt{\pi}}\frac{v_c}{\vt}e^{-v_c^2/\vt^2}\bigg](1-\cos\theta_c).\nonumber
\end{align}
From this expression it is clear that the amount by which the electron density is decreased is a function of $v_c/\vt=\sqrt{|e\Delta\phi|/T_e}$. 

\begin{figure}[h]
    \centering
    \includegraphics[width=0.45\textwidth]{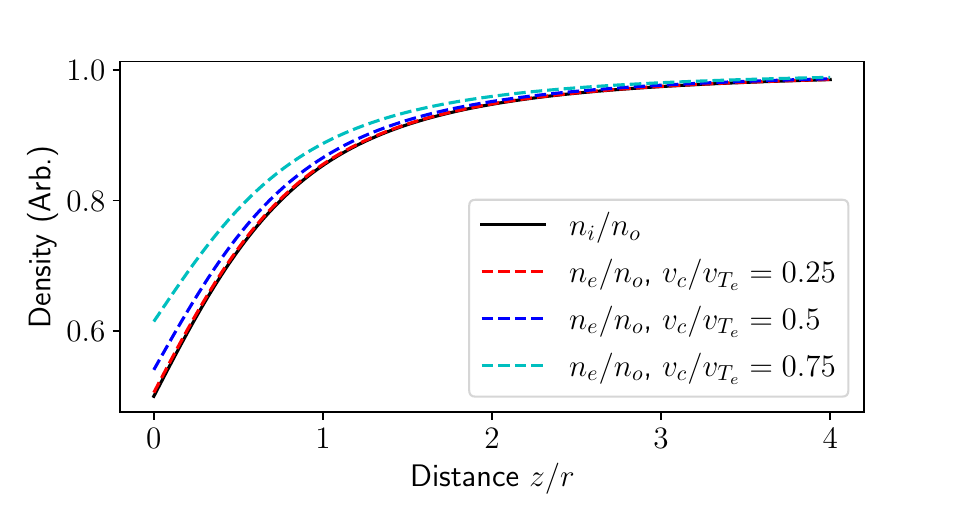}
    \caption{Comparison of the electron and ion density profiles for the case where a weak electron retarding potential is applied to the electrode. The cases of $v_c/\vt$ = 0.25, 0.5, and 0.75 are shown. These correspond to electrode biases of $T_e/8e$, $T_e/4e$, and $3T_e/8e$ below the plasma potential, respectively.}
    \label{fig:belowpp}
\end{figure}

Figure~\ref{fig:belowpp} shows a comparison of profiles of $n_e$ for $|\Delta\phi_R|=T_e/8$, $T_e/4$, and $3T_e/8$ with the ion density profile given by Eq.~\ref{eq:ne}: $n_i/n_o=0.5(1+\cos\theta_c)$. 
%(assuming the loss cone distribution is a good approximation for small biases below the plasma potential). 
The most notable feature of the comparison is that for retarding potentials of $|\Delta\phi_R|\lesssim T_e/4$ the electron and ion densities are nearly equal and that quasineutrality can be maintained up to the electrode. 
This suggests the absence of an ion sheath at the electrode for these conditions. The absence of an ion sheath or presheath electric field justifies the use of the free streaming trajectory assumption.    

Now consider the analogous situation with the electrode biased positive with a slight retarding potential for ions. 
For this case, the electron and ion subscripts on the density equations, thermal speeds, and mass in $v_c$ are interchanged. The conclusion is similar: if the potential is not more than $\sim T_i/4$ above the plasma potential quasineutrality can be maintained. 
Since $T_e\gg T_i$ in most laboratory low temperature plasma experiments, there is asymmetric behavior for positive and negative biases about the plasma potential. 
If for example we consider an electron temperature of $T_e\approx2$ eV and an ion temperature of $T_i = 0.08$ eV, a bias of -0.25V below the plasma potential results in the absence of a sheath, while a bias 0.25 above the plasma potential is sufficient for the formation of an electron sheath. This behavior has been confirmed in PIC simulations with conditions very close to those used in this example\cite{10.1063/1.4960382}.

\subsection{Electron Presheath\label{sec:epre}}
In this subsection the behavior of the electron presheath is considered. 
As mentioned in the introduction, the electron presheath is an extended quasineutral region outside of an electron sheath in which the plasma density and electron flow velocity are perturbed from their bulk plasma values. 
Here the predicted properties of the electron presheath are based on the ansatz that the electrons are collected at the electron sheath edge by the same effusion process as in the case of a probe biased at the plasma potential (Sec.~\ref{sec:pp}). 

\begin{figure}[h]
    \centering
    \includegraphics[width=0.4\textwidth]{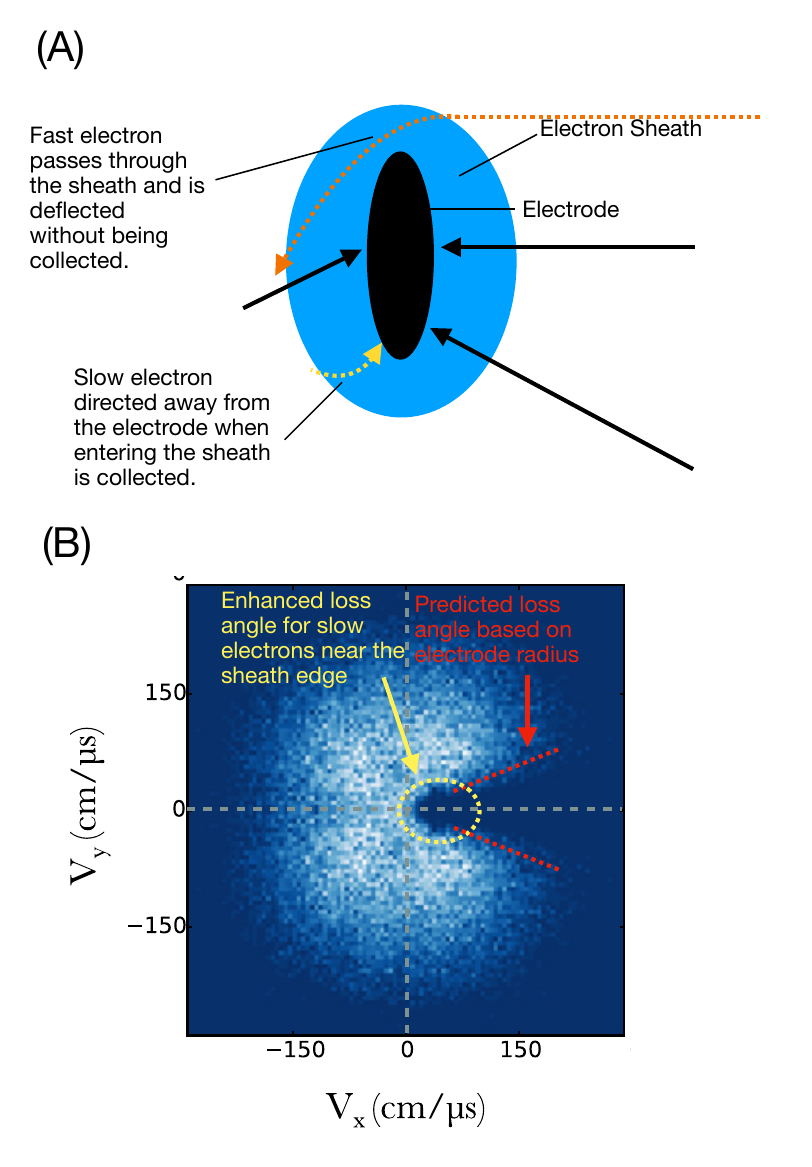}
    \caption{(A) Examples of electron trajectories through an electron sheath. Fast electrons passing through the sheath that are not directed at the electrode surface (red) are not be collected. Slow electrons passing the sheath edge (yellow) are collected regardless of their direction.  (B) Examples of the impacts of the fast and slow electron populations on the presheath EVDF. Adapted from B. Scheiner et al. Phys. Plasmas 23, 083510 (2016), with the permission of AIP Publishing.}
    \label{fig:EsheathTraj}
\end{figure}

As shown in Fig.~\ref{fig:EsheathTraj}(A), there is the complication that not all electrons passing the sheath edge are collected by the electrode. Generally, whether or not an electron crossing the sheath edge is collected is a function of its energy, direction, and location relative to the electrode. For example, those crossing the sheath edge near the electrode center are very likely to be collected, while those crossing the sheath edge at a radial distance past the edge of the electrode have an enhanced likelihood of passing through. Likewise, slow electrons passing the sheath edge at a radial location past the edge of the electrode are likely to be collected while fast electrons may not be depending on the orientation of their velocity vector. Figure~\ref{fig:EsheathTraj}(B) demonstrates a realization of these effects in an annotated EVDF extracted 0.25 cm from a $r=0.1$ cm electrode in the PIC simulation of Ref.~\onlinecite{10.1063/1.4960382}. The geometric loss cone angle marked by the red line is enhanced at slow velocities. As indicated by the yellow circle, enhanced collection occurs for slow electrons. 
These effects act to enhance the effusion loss area for electrons from the bulk plasma. 
Furthermore, electron deflection by the sheath acts to enhance the positive velocity portions of the EVDF that are just outside of the depletion region. 
In the limit that the sheath width is small compared to the electrode radius, neither effect is important and the effective value of $r$ in the calculation of $\theta_c$ is the electrode radius. 
Estimates of the sheath thickness can be made using the Child-Langmuir law for electron sheaths\cite{10.1063/1.1887189}. 
%Neglecting the effect of orbits, once an electron intercepts the electrode it is collected by the electrode. For the disk electrode geometry considered in this paper orbits are unlikely, but may need to be considered for cylindrical or spherical probes. With this in mind, 
The presheath properties can be predicted using the theory of Sec.~\ref{sec:pp} once the disk radius $r$ is replaced by an effective sheath-modified radius $r_{\textrm{eff}}$. A more detailed theory may treat high and low velocity electrons with different loss angles, but such treatment is beyond the scope of this work.

\begin{figure}[h]
    \centering
    \includegraphics[width=0.45\textwidth]{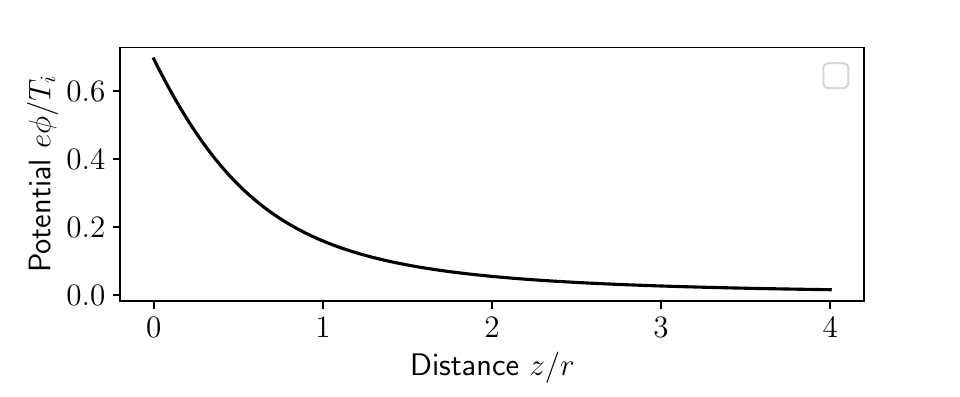}
    \caption{Quasineutral plasma solutions for the electron presheath potential profile. Electrode biases more than $\sim0.7T_i/e$ above the plasma potential result in sheath formation.}
    \label{fig:phi}
\end{figure}

Another point that has gone unmentioned thus far is the value of the electrostatic potential in the electron presheath. 
If a presheath potential drop of magnitude $T_e/2$ were present as it is in the ion presheath the free streaming approximation of electron trajectories would break down. 
However, previous simulations have suggested that the electron presheath electric field is negligible. In fact, the theory provided in this work is consistent with this observation.
The potential profile in the presheath can be obtained by assuming a Boltzmann relation for ions $n_i=n_o\exp(-e\phi/T_i)$ and
equating the electron density from Eq.~\ref{eq:ne} with ion density in the presheath to solve for $\phi$. 
The potential profile along the axis perpendicular to the electrode center is
\begin{equation}\label{eq:pp}
    \phi=-\frac{T_i}{e}\ln\big[\frac{1}{2}(1+\cos\theta_c)\big].
\end{equation}
The plot of this profile in Fig.~\ref{fig:phi} indicates that the potential varies by about $T_i\ln(2)/e$ between the bulk plasma and the absorption surface. 
%The actual presheath potential at the sheath edge is somewhat less than this value since the quasineutral solution breaks down within the sheath before reaching the electrode surface at $z=0$. 
For plasmas with $T_i<<T_e$, a potential drop of $T_i\ln(2)/e$ is not enough to appreciably change the electron trajectories in the presheath. When $T_i$ becomes comparable to $T_e$ the presheath electric field will modify the electron trajectories resulting in a breakdown of the free streaming trajectory approximation.          

\subsubsection{Breakdown of quasineutrality}

Here, we briefly consider the breakdown of quasineutrality resulting in the formation of an electron sheath. Expanding the charge density about a point at the sheath edge with potential $\phi_0$ above the plasma potential, $\rho(\phi)=\rho(\phi_0)+|d\rho/d\phi|_{\phi=\phi_0}(\phi-\phi_0)+...$. At this point $\rho(\phi_0)=0$, but  $|d\rho/d\phi|_{\phi=\phi_0}<0$. This requirement is known as the sheath criterion\cite{K-URiemann_1991} and can be restated as
\begin{equation}
    \sum q_s\frac{dn_{s}}{dz}\le0. 
\end{equation}
Inserting the derivative of the Boltzmann relation for ions 
\begin{equation}
    \frac{dn_i}{dz}=-\frac{en_i}{T_i}\frac{d\phi}{dz}=\frac{en_i}{T_i}E
\end{equation}
the derivative of Eq.~\ref{eq:ne} for electrons
\begin{equation}
    \frac{dn_e}{dz}=-\frac{n_o\sin\theta_c}{2}\frac{d\theta_c}{dz},
\end{equation}
and making use of the quasineutrality relation $n_e=n_i$ with Eq.~\ref{eq:ne} results in 
\begin{equation}
E\le-\frac{T_i}{e}\tan(\theta_c/2)\frac{d\theta_c}{dz}.
\end{equation}
This value is exactly the value obtained by taking the derivative of the presheath potential profile in Eq.~\ref{eq:pp}. Sheath formation results from the presence of an electric field (potential) greater than that of the profile Eq.~\ref{eq:pp}. The maximal probe bias before electron sheath formation is $\phi=-T_i\ln(1/2)/e$.

\section{Comparison with prior simulation data}
This section tests the predictions of Sec.~\ref{sec:pp} and ~\ref{sec:epre} against existing PIC simulation data from Ref.~\onlinecite{10.1063/1.4960382} which were computed using 2D-3V simulations. 
To compare with simulations, a 2D version of the theory of Sec. II is needed. In 2D PIC, particles in the plane can be viewed as rods with arbitrary out of plane dimension. 
Therefore, for comparison between theory and simulation, the distribution function and its moments should be defined and computed over their values in 2D since the velocities in the out of plane direction are not affected by the presence of the boundary. The 2D VDF is
\begin{equation}\label{eq:2dvdf}
f_{e,2D}=\frac{ n_o}{\pi\vt^2 }\exp\big(-v^2/\vt^2\big)\big[H\big(\theta-\theta_c\big)+H\big(-\theta_c-\theta\big)]
\end{equation}
where $\theta\in[-\pi,\pi]$ is aligned along the electrode normal. The normalization was determined such that the density moment 
\begin{equation}
    n_e=\int_0^\infty \int_{-\pi}^{\pi} f_e(v,\theta) v dv d\theta
\end{equation}
yields $n_o$ when $\theta_c\to 0$. Using this definition, the 2D density profile is 
\begin{equation}\label{eq:ne2D}
n_e=\frac{n_o}{\pi}(\pi-\theta_c).
\end{equation}
Likewise, the 2D z flow velocity is 
\begin{equation}
    V_{e,z}=\frac{1}{n_e}\int_0^\infty \int_{-\pi}^{\pi} f_e(v,\theta) v \cos\theta v dv d\theta,
\end{equation}
which gives
%\begin{equation}
%    V_{e,z}=\frac{\pi}{n_o(\pi-\theta_c)}[\frac{ n_o}{\pi\vt^2 }\frac{\sqrt{\pi}\vt^3}{4}][-2\sin\theta_c]
%\end{equation}
\begin{equation}\label{eq:ve2D}
    V_{e,z}=-\frac{\sqrt{\pi}\vt}{2}\frac{\sin\theta_c}{(\pi-\theta_c)}.
\end{equation}
The electrode in the simulations of Ref.~\onlinecite{10.1063/1.4960382} is a line segment with length $r=0.1$ cm. For this the angle subtended by the electrode at distance z is unchanged and the definition of $\theta_c$ from Sec.~\ref{sec:pp} applies. 

\subsection{Electrode biased at the plasma potential\label{sec:pppic}}
\begin{figure}[h!]
    \centering
    \includegraphics[width=0.4\textwidth]{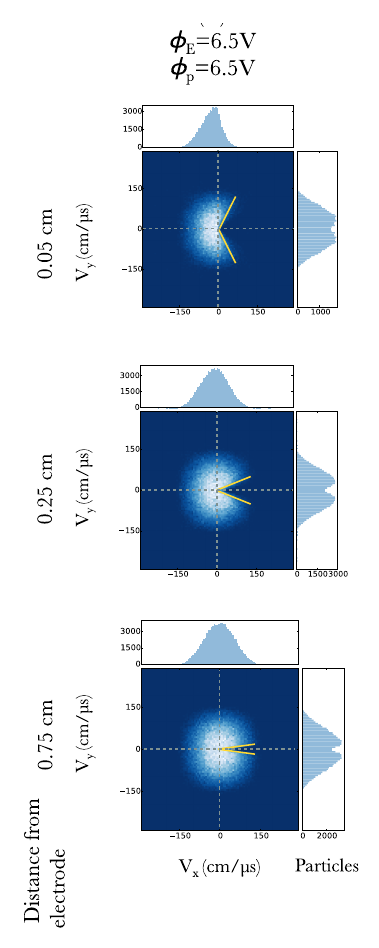}
    \caption{Comparison of the predicted EVDF loss cone angle (yellow lines) against the depletion region of the EVDF obtained from PIC simulations. Adapted from B. Scheiner et al. Phys. Plasmas 23, 083510 (2016), with the permission of AIP Publishing.}
    \label{fig:NoSheathCone}
\end{figure}

\begin{figure}[h]
    \centering
    \includegraphics[width=0.45\textwidth]{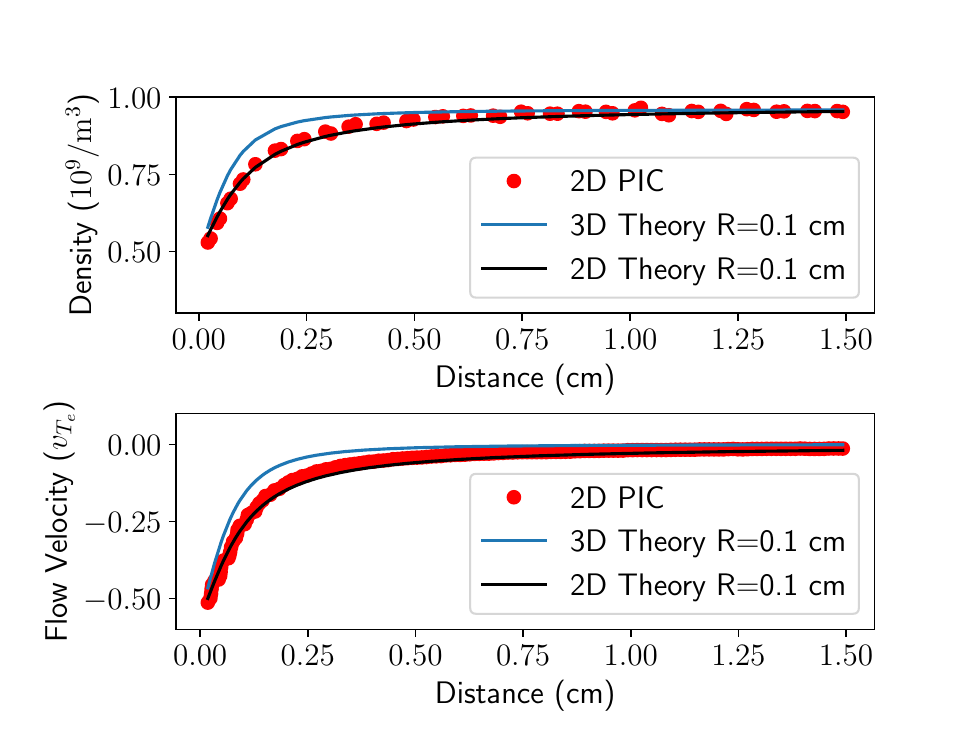}
    \caption{Comparison of 2D theory (Eqs.~\ref{eq:ne2D} and \ref{eq:ve2D}) and 3D theory (Eqs.~\ref{eq:ne} and \ref{eq:ve}) with PIC data for axial flow and density profiles extracted from Ref.~\onlinecite{10.1063/1.4960382}.}
    \label{fig:NoSheathComp}
\end{figure}

Here we consider the case of the electrode biased at the plasma potential.
The theory predicts (1) the loss cone angle of the EVDF and (2) the velocity and density profiles. Starting with (1), we compare the loss cone angle of the EVDF obtained from PIC simulations to the prediction of the angle given by Eq.~\ref{eq:thetac}.
In Ref.~\onlinecite{10.1063/1.4960382}, the EVDF was computed at axial distances of 0.05, 0.25, and 0.75 cm from the electrode. For an electrode of diameter 0.2cm, the cone angle for these distances are $\theta_c =$ 63.4$^o$, 21.8$^o$, and 7.6$^o$, respectively. 
These angles are drawn on the EVDFs in Fig.~\ref{fig:NoSheathCone} and are in agreement with the boundary of the depleted region of the distribution. 
Figure~\ref{fig:NoSheathComp} compares the density and flow velocity profiles from the 2D (Eqs.~\ref{eq:ne2D} and \ref{eq:ve2D}) and 3D (Eqs.~\ref{eq:ne} and \ref{eq:ve}) theory with profiles from PIC\footnote{Note that the profiles in Ref. 9 were incorrectly normalized by $\vt=\sqrt{kT_e/m_e}$ instead of $\vt=\sqrt{2kT_e/m_e}$. The correction factor has been applied to the PIC data. The original data from this study is no longer available and data points were extracted from published figures.}.
The 2D theory provides an excellent description of the simulation data. The 3D theory predicts more shallow profiles and are shown to emphasize the expected difference between 2D and 3D simulations of the same phenomenon.  

%Next the profiles of the 2D and 3D theory are compared with the profiles from simulation in Fig.~\ref{fig:NoSheathComp}. 

\subsection{Electron presheath}

\begin{figure}[h]
    \centering
    \includegraphics[width=0.4\textwidth]{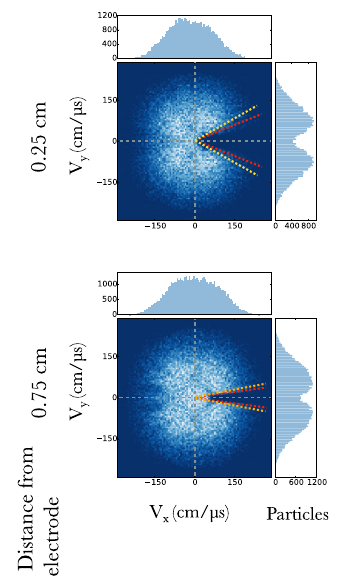}
    \caption{Comparison of the electron sheath loss cone angles for an effective electrode area of $r=0.1$ cm (red) and $r=0.14$ cm (yellow) at two different locations in the presheath. Adapted from B. Scheiner et al. Phys. Plasmas 23, 083510 (2016), with the permission of AIP Publishing.}
    \label{fig:EsheathCone}
\end{figure}

Similar to Sec.~\ref{sec:pppic}, this section makes comparisons for the case of an electron sheath. In the electron sheath simulation, the electrode potential was 25V, the plasma potential was 18.5V, and the sheath edge in the axial direction was approximately 0.18 cm (defined by more than $\sim$5\% deviation from quasineutrality). 
First we compare the angle of the loss cone in the presheath with the depletion in the calculated EVDF in Fig.~\ref{fig:EsheathCone}. The EVDF was calculated at the same positions as in Sec.~\ref{sec:pppic}, 0.05, 0.25, and 0.75 cm. For the electron sheath simulations only the later two are in the presheath. In the comparison with simulation data we consider the possibility suggested in Sec.~\ref{sec:epre} that the thickness of the sheath at the electrode enhances the radius at which electrons are collected. 
For this purpose we consider an effective electrode loss radius of $r=0.1$ (unmodified), 0.12, and 0.14 cm. The larger and smaller angle corresponding to $r=0.14$ cm and 0.1 cm, respectively, are shown and are 21.8 and 29.3 degrees at 0.25 cm and 7.6 and 10.6 degrees at 0.75 cm. Because the larger angle includes some high velocity regions of the EVDF that are not entirely depleted, the angle corresponding to the smaller electrode radius is a better fit for larger electron velocities. As expected from the discussion in Sec.~\ref{sec:epre}, these electrons have a higher likelihood of passing through the sheath without being collected by the electrode and do not experience enhanced collection. 

\begin{figure}[h]
    \centering
    \includegraphics[width=0.45\textwidth]{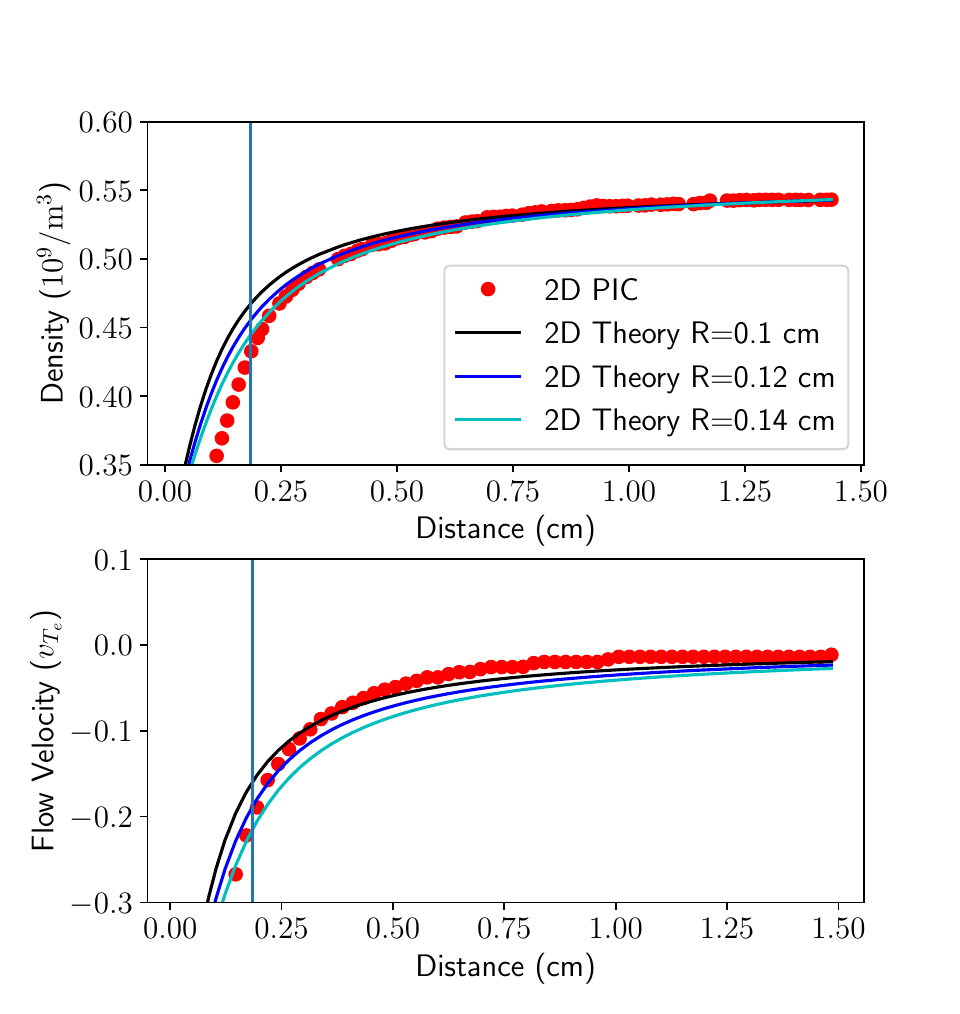}
    \caption{Compares presehath density and flow velocity profiles with PIC for the cases of an effective electrode loss radius of $r=0.1$, 0.12, and 0.14 cm. The vertical line indicates the location of the sheath edge.}
    \label{fig:Esheath}
\end{figure}

Figure~\ref{fig:Esheath} compares presehath density and flow velocity profiles with PIC for the cases of an effective electrode loss radius of $r=0.1$, 0.12, and 0.14 cm. Irrespective of the precise value of the electrode radius, the model does a good job of predicting the length scale of profiles seen in the presheath. For the density, small deviations are seen between 0.25 cm and the sheath edge location of 0.18 cm indicated by the vertical line in the figure. The flow velocity profile with $r=0.1$ cm agrees best with the data, but slightly overestimates the PIC values, being within 50\% at 0.8 cm and 20\% at 0.5 cm. Finally, the simulations exhibited little change in the electrostatic potential within the presheath, and the value of the electric field term in the previous momentum equation analysis of Ref.~\onlinecite{10.1063/1.4960382} was minimal. This is in agreement with the weak potential gradient suggested by Eq.~\ref{eq:pp}.  

 It is interesting to note that the effects of electron deflection can be observed in the EVDF computed within the sheath in Ref.~\onlinecite{10.1063/1.4960382}. At locations near and just past the sheath edge, the high velocity regions of the EVDF with positive velocity are populated and the cone angle does not fully reach the 90 degree value at the sheath edge. Inclusion of the effects of a velocity dependent loss cone and electron deflection in the sheath into the model is expected to improve agreement.

\section{Discussion}

\subsection{Plasma analog of a Knudsen gas}
A Knudsen gas\cite{knudsen} is defined as the regime in which particles undergo free molecular flow, colliding more frequently with boundaries than with each other. The extent to which a gas exhibits this behavior is characterized by the Knudsen number $\textrm{Kn}=\lambda_{mfp}/L$, where $L$ is the scale of the gas container and $\lambda_{mfp}$ is the collisional mean free path. A Knudsen gas satisfies $\textrm{Kn}\gg1$, while a typical fluid has $\textrm{Kn}\ll1$.
In this paper a theory for the plasma perturbation posed by probes biased near and above the plasma potential in collisionless plasmas was developed on the basis of the free streaming particle or free molecular flow approximation and effusion loss at a conducting electrode. Contrasting with the case of the ion presheath where ions are accelerated to the Bohm speed by a $T_e/2e$ presheath potential\cite{10.1063/1.1887189}, the collisionless electron presheath properties are determined only by the loss area of plasma boundaries and electrons do not change velocity otherwise outside of the boundary region. Its properties are determined entirely non-locally by the boundary. In this sense electrons in the vicinity of electrode are a realization of a Knudsen gas. 

One limitation of the Knudsen gas analogy is the volumetric source rate of electron in the plasma.
%The first is that the description does not hold for ions since their trajectories are affected by the weak $\sim T_i/e$ presheath potential. 
%A second limitation concerns the source of particles in the plasma. 
In common laboratory DC discharge plasmas, such as those in multi-dipole chambers, electrons and ions are formed at a near constant rate in the chamber volume by electron impact ionization from an energetic primary electron population. 
For the Knudsen gas description to hold, the phase space of electrons in the vicinity of the electrode should have behavior determined by the collection from the electrode (e.g. absence of certain trajectories due to the formation of the loss cone). If the source rate is too great, the loss cone will fill due to the isotropic sourcing of electrons in the volume.
In practice, this may limit the Knudsen gas description to regions within a distance $L^\prime$ of the boundary. An approximate criteria for this limitation is the ratio of the ionization frequency $\nu_{iz}$ to the average inverse transit time of an electron to the boundary $\bar{v}_e/L^\prime$, where $\bar{v}_e=\sqrt{8kT_e/\pi m_e}$ is the mean electron speed. Combined with the Knudsen number requirement, the validity of the Knudsen gas description of electrons requires 
\begin{equation}\label{eq:condition}
    \frac{\bar{v}_e}{\nu_{iz}L^\prime}\gg1, \ \ \ \textrm{Kn}\gg 1.
\end{equation}

An additional limitation of the Knudsen gas description has to do with the growth of kinetic instabilities. Many configurations of the EVDF are possible depending on the electrode geometry. Unstable configurations may result in the growth of waves which may be accompanied by sufficient variations in potential such that the scattering and trapping of electrons may occur. Such effects may result in the change in momenta of electrons in the vicinity of the electrode such that the electron mean free path is shorter than the length scale of the plasma device.  

Another related consideration is the anomalous thermalization of the electron VDF known as Langmuir's Paradox\cite{PhysRev.26.585,gabor1955langmuir}.
Langmuir measured a sustained high energy tail in the EVDF even though these electrons had sufficient energy to overcome the sheath potential at the bounding walls of his experiment. The thermalization was unexpected since the collisional mean free path was 5-10 times longer than that which would be needed to explain the observation\cite{PhysRev.26.585}. Different solutions have been put forward, including the electron interaction with standing striations
\cite{Godyak_2015} and an enhanced collision rate for electrons in the presence of instabilities\cite{PhysRevLett.102.245005}. It should be noted that the PIC simulations are capable of modeling an instability enhanced collision rate\cite{10.1063/1.5089507} and that thermalization was not observed in the simulations that were used to test the theory.

\subsection{Comparison with the previous collisional electron presheath fluid model}
The present theory has several similarities with the collisional presheath model in Ref.~\onlinecite{10.1063/1.4939024}. In this section we contrast the features of the two. In the previous work, the presheath was described by a modified mobility limited flow equation obtained by assuming a Boltzmann relation for ions and inserting a quasineutrality condition into the continuity and momentum equations:  
\begin{equation}
    V_e=-\mu_e\bigg(1-\frac{V_e^2}{v_{eB}^2}\bigg) E.
\end{equation}
Here, E is the electric field, $\mu_e=e(1+T_e/T_i)/[m_e(2\nu_{c}+\nu_{iz})]$ is the electron mobility determined by the collision and source ionization frequencies, and  $v_{eB}=\sqrt{T_e+T_i/m_e}$ is the electron Bohm speed. The key features of this model were:
\begin{enumerate}
    \item The presence of presheath electron flows with velocities approaching the electron thermal speed.
    \item A presheath length scale determined by the collisionality or source rate $l_e=v_{eB}/\nu$ that is much longer that the corresponding ion presheath model ($l_i=c_s/\nu$) under similar conditions.
    \item The presence of pressure gradient driven electron flows.
    \item Weak potential gradients with the potential varying by $\sim T_i/e$ over the presheath length.
\end{enumerate}
In light of the work presented in this paper, such a model is appropriate when either of the quantities from Eq.~\ref{eq:condition} are much less than unity.   

For the collisionless theory several of the key features from the collisional model remain. A flow velocity of $\sim0.1-0.2\sqrt{2kT_e/m_e}$ is observed in the presheath, roughly 20-30\% of the values expected from the collisional theory. In the collisionless case this velocity increases over the length of the presheath which is determined by the size of the electrode, but still may extend a significant distance into the plasma. This length scale is not necessarily longer than that for an ion presheath under similar conditions. When the collisional mean free path is longer than the chamber scale (i.e. $\textrm{Kn}>1$), the ion presheath length becomes half the plasma chamber length\cite{4316372}. For both models, minimal potential gradients are present in the presheath due to the ease at which ions are repelled from the boundary, owing to their much lower temperature. Finally, a fluid moment analysis of each theory would indicate the dominance of the pressure gradient term over the potential gradient term in an analysis of the electron momentum balance. In the collisionless case this is purely a non-local effect due to the geometry of the electrode. For the model distribution function in Eq.~\ref{eq:vdf}, the derivative of the $\hat{z}\hat{z}$ term of the pressure tensor varies spatially as 
%\begin{equation}
%    P_{e,zz}=\frac{3}{4}n_o\vt^2m_e(1+\cos\theta_c)
%\end{equation}
\begin{equation}
    \frac{dP_{e,zz}}{dz}=\frac{3}{4}n_o\vt^2m_e\frac{r^2}{(r^2+z^2)^{3/2}}.
\end{equation}
This is $\sim(T_e/T_i)(1+\cos\theta_c)$ larger than the corresponding electric field term in the momentum equation calculated from Eq.~\ref{eq:pp}.

For plasmas with $\textrm{Kn}\sim 1$ the results of neither theory strictly apply. This behavior is analogous to the transitional flow regime in the rarefied gas dynamics literature. This regime is the most difficult to describe because the boundary will induce some amount of non-local behavior in the flow. For these cases, elements in common to both collisional and collisionless theories are expected. A description of this behavior is an open problem and is beyond the scope of the present work. 
%Key differences are: (1) the collisional theory predicts a pressure gradient driven flow towards the boundary. This is the correct picture for the case where the boundary is larger than the collisional mean free path. (note transition in effusion to bulk flow for neutral gas case). (2) For the collisional model the quasineutral solution diverges at the electron bohm speed. In the collisionless theory, the breakdown of the quasineutral solution occurs because quasineutral solutions cannot be maintained for potentials greater than $\sim T_i/e$. (3) In the collisional theory, the presheath length scale is determined by the collisional mean free path while in the collisionless theory it is determined by the electrode scale. 

%In both cases presheath pressure gradients are expected to be larger than the electric field gradient as both theories predict this is zero. In the collisionless theory, the pressure gradient arises solely due to the non-local modification of the evfd. 

\subsection{Electron Sheath Bohm Criterion}
Prior to the observation of an electron presheath\cite{Yee_2017,10.1063/1.4939024,10.1063/1.4960382}, the conventional wisdom was that the electron flux collected by the electrode is the random flux \cite{10.1063/1.1887189, PhysRev.28.727} and that as a result the electron sheath Bohm criterion is trivially satisfied\cite{K-URiemann_1991,Chen_2006}. Implicit in these statments is that the EVDF at the sheath edge is a half-Maxwellian. The present theory predicts such a half-Maxwellian EVDF and is consistent with the conventional expectations of the collection of the random flux, a result which has been indicated by the recent low pressure experimental observations in Ref.~\onlinecite{Jin_2022}. The presheath predicted in this paper explains how the half-Maxwellian EVDF is generated by the sheath edge and how a presheath can still be present in the absence of a Bohm criterion. The previous prediction of an electron sheath Bohm criterion\cite{Yee_2017,10.1063/1.4939024,10.1063/1.4960382} applies only to situations where the electron mean free path is much less than the presheath length.

\section{Conclusion}

This paper presents a new theory for an effusion based loss mechanism of electrons for probes biased at electron saturation and near the electron saturation transition. The theory predicts the existence of a presheath with a very weak electric field whose properties are determined non-locally by the electrode geometry and the loss of particles whose free streaming trajectories intercept the conducting boundary. The results were compared with previously published PIC simulations and are in good agreement with density and flow profiles and the predicted EVDF is in good agreement with those from the simulation. Slight deviations that can be attributed to particles passing through the sheath uncollected were observed for the case of the electron presheath. 
An analogy was drawn between the electrons in the vicinity of the electrode and a Knudsen gas and this comparison was used to demonstrate that the theory requires $\textrm{Kn}\gg1$ and weak volume ionization in the gas. 
The theory was contrasted with the previous collisional electron presheath theory and the range of validity of the collisional and collisionless theory was provided to clarify the situations in which each should be applied. Finally, the theory was found to be consistent with recent low pressure experimental measurements of Ref.~\onlinecite{Jin_2022} and the conventional wisdom that electron sheath collects a random flux of electrons.      

\section*{Acknowledgements}
The author thanks Louis Jose, Lucas Beving, and Scott Baalrud for their comments on the manuscript. The author also thanks Scott Baalrud for his suggestion of the effusion vs diffusion analogy. 

\section*{Data Availability}
The data that supports the conclusions of this study are contained in the manuscript. 

\bibliography{apssamp}
\end{document}